\documentclass[aps,pre,twocolumn,superscriptaddress,amsfonts,floatfix]{revtex4-1}
\usepackage{xcolor,todonotes,layouts}

\begin{document}

 \title{Stable and Unstable Vortex Knots in Excitable Media}
 \author{Jack Binysh}
 \affiliation{Mathematics Institute, Zeeman Building, University of Warwick, Coventry, CV4 7AL, United Kingdom}
  \author{ Carl A. Whitfield}
  \affiliation{Division of Infection, Immunity and Respiratory Medicine, University of Manchester, Southmoor Road, Manchester, M23 9LT, United Kingdom }
   \author{Gareth P. Alexander}
 \affiliation{Department of Physics and Centre for Complexity Science, University of Warwick, Coventry, CV4 7AL, United Kingdom}
\date{\today}

\begin{abstract}
    { We study the dynamics of knotted vortices in a bulk excitable medium using the FitzHugh-Nagumo model. From a systematic survey of all knots of at most eight crossings we establish that the generic behaviour is of unsteady, irregular dynamics, with prolonged periods of expansion of parts of the vortex. The mechanism for the length expansion is a long-range `wave slapping' interaction, analogous to that responsible for the annihilation of small vortex rings by larger ones. We also show that there are stable vortex geometries for certain knots; in addition to the unknot, trefoil and figure eight knots reported previously, we have found stable examples of the Whitehead link and $6_2$ knot. We give a thorough characterisation of their geometry and steady state motion. For the unknot, trefoil and figure eight knots we greatly expand previous evidence that FitzHugh-Nagumo dynamics untangles initially complex geometries while preserving topology.} 
\end{abstract}
\pacs{} 

 \maketitle
 
\section{Introduction}

Models of excitable media support spiral wave vortices in two dimensions. In a three-dimensional medium the analogous structure is a vortex filament \cite{Winfree1983}. Such a filament may close on itself to form, in the simplest case, an unknotted loop, and more generally a knotted vortex \cite{Winfree1983b}. As well as being organising centres for waves of excitable activity, early numerical experiments and theoretical work showed that these knotted filaments have their own dynamics \cite{Keener1988,Winfree1990, Keener1992,Henze1993, Biktashev1994,WinfreeChapter,Dierckx2010}. Remarkably, in a simple example of an excitable medium, the FitzHugh-Nagumo model, simulations suggested that these dynamics were topology preserving, and further that they were capable of `simplifying' a knot, reducing an initially complicated filament geometry to a simpler stationary state~\cite{Winfree1990,Henze1993,WinfreeChapter}. Such a scenario stands in stark contrast to the knot untying via reconnection events seen in other examples of knotted fields such as fluids and superfluids \cite{Kleckner2013,Scheeler2014,Kleckner2016}. 

So far, the striking knot simplification has been reported only for the unknot \cite{Maucher2016} --- reducing three examples of tangled, but unknotted, curves to a geometric circle of fixed average radius --- and, in two examples, the trefoil~\cite{Sutcliffe2003,Maucher2016}. At higher crossing number the behaviour appears to be more complicated. Nevertheless, with the exception of two examples discussed below, reported knot and link evolutions are still consistent with preservation of topology \cite{Henze1993, Winfree1990, WinfreeChapter, Sutcliffe2003,Maucher2016,Maucher2017, Maucher2018b}. Stationary states have been reported for all torus knots and links up to $N=12$~\cite{Maucher2017,Maucher2018b}, where $N$ denotes the crossing number of the knot or link, although only in the case where they are stabilised by proximity to a planar surface with Neumann (no-flux) boundary conditions and the vortex filaments are initialised to have the idealised geometry of torus curves. Both of these features are believed to be important for the stability of these examples \cite{Maucher2017,Maucher2018b}. If the vortex is initialised with idealised torus geometry but not sufficiently close to the boundary, it is prone to poorly understood instabilities which cause it to deviate from the initially symmetric form. Similarly, torus knots initialised without the idealised geometry do not evolve to the observed stationary states and instead follow irregular dynamics, typically ending with the filament breaking upon contact with the no-flux surface. It is not known how close to the idealised torus geometry the initial curve needs to be to attain the stationary state. A cautionary example is provided by early bulk (not close to a no-flux boundary) simulations of a variety of knots and links, including torus knots, started from exactly symmetric configurations, which appeared stable over the times initially simulated~\cite{Henze1993}. As we shall demonstrate in the case of torus knots, over longer timescales such geometries in fact destabilise. The recent no-flux simulations are over much longer timescales, however in the absence of theoretical results it is not clear exactly how long a simulation is `long enough'.

As indicated above, in contrast to topology preserving dynamics seen in the unknot some apparently exceptional examples of topological non-preservation have been reported, the first being the already discussed example of a single filament breaking at a no-flux boundary, effectively interacting with its mirrored neighbour. In the bulk, a small vortex loop (stable in isolation) may be annihilated by a larger coaxial one ~\cite{Courtemanche1990,Maucher2018a}. The mechanism behind this annihilation is thought to be `wave slapping' by a train of high frequency wavefronts coming from the larger loop impacting upon the smaller one (which has a vortex rotation frequency below that seen for an isolated filament) causing it to destabilise. 

Underlying all of the above observations are general questions as to the driving factors behind knot dynamics, which can be expected to involve both the topology of the vortex and its geometry. 
Theoretical work has focused on local geometric models of filament motion in which an isolated straight filament is perturbed to have slight curvature and longitudinal twist in the phase of its cross sectional spiral waves~\cite{Keener1988,Keener1992,Henry2002,Biktashev1994,Echebarria2006,Dierckx2010}. To lowest order these deformations lead to filament motion with both normal and binormal components, as well as a modification of vortex rotation frequency away from the intrinsic (straight filament) frequency of the excitable medium~\cite{Keener1988,Biktashev1994,Dierckx2010}. A `sproing' instability, in which an initially straight filament twisted above some critical threshold destabilises and adopts a helical conformation, has also been predicted and observed~\cite{Keener1988,Henze1993,Henry2002,Echebarria2006, Dierckx2010}, which has been proposed~\cite{Maucher2017} to account for the deformations seen in torus knot simulations. 

These considerations are all local and do not capture the global topology of the vortex or non-local wave-vortex interactions, which are essential to the process of simplification without reconnection observed for unknotted loops. The nonlinear waves of excitation propagating from the vortex filament mutually annihilate when they meet, creating a complex `collision interface'~\cite{Henze1991, Winfree1990, Henze1993, WinfreeChapter, Sutcliffe2003} depending not only on the filament geometry but also on the synchrony of wave emission from distant parts of the vortex (Ref.~\cite{Henze1991} shows an example of this interface for a trefoil knot with a different kinetics to that considered here). The analogous structure in a two-dimensional medium, sometimes known as a ``shock structure"~\cite{Kopell1973}, is well studied in a variety of excitable media \cite{Krinsky1983,Ermakova1986,Vinson1998,Gottwald2001,Agladze2007,Steinbock2011}. If a two-dimensional spiral vortex interacts with a wavefield of frequency higher than its own (as generated either by other vortices or externally) this interface moves towards the low frequency vortex until directly upon it, at which point the high frequency wavefield directly slaps the vortex. This slapping may then induce motion in the vortex, commonly referred to as ``spiral wave drift" or ``high frequency induced drift". The same scenario may occur in three dimensions --- a more general instance of the wave slapping mechanism described above for a pair of coaxial rings --- and has been proposed as an important driver of filament dynamics, a conjecture for which there is some indirect evidence \cite{WinfreeChapter, Sutcliffe2003,Maucher2018b}. However, in the three-dimensional case the factors that determine the collision interface are poorly understood --- as discussed above, local filament geometry may in principle affect vortex rotation period, but the observed frequency shift and annihilation of a small unknot suggests that inter-filament interactions and Doppler shift due to relative filament motion are also important.

We present here the results of a systematic survey of the dynamics of all prime knots up to crossing number $N=8$ and focus on behaviour in the bulk, using periodic boundary conditions, so as to further complement recent work by Maucher \& Sutcliffe~\cite{Maucher2016,Maucher2017,Maucher2018a,Maucher2018b}, who have studied no-flux boundary conditions. We find generically that knotted vortices do not stabilise into simplified stationary states, although some do. The predominant behaviour is of unsteady dynamics and instability through the expansion of some portion of the knot into a large loop; we present evidence that this instability occurs through the wave slapping mechanism alluded to previously. In a substantial fraction of cases (eight of thirty-six) the instability eventually leads to strand reconnections in the bulk, demonstrating that such events are in fact not exceptional as one increases crossing number, and do not occur solely at a boundary or in a highly symmetric geometry. The reconnections are of anti-parallel strands, driven together through wave slapping in a manner analogous to the annihilation of the unknot discussed above, and result in links. For both a generic knot and the specific case of idealised torus knots we additionally investigate the role of the sproing instability in knot destabilisation, finding it to be unimportant in explaining generic knot instability and not directly responsible for torus knot destabilisation, although correlated with torus knots attaining a temporarily stable knot length.

Our survey also shows that for $N\leq4$ (unknot, trefoil, figure eight) knots do exhibit topology preserving dynamics towards stationary states. We strengthen these results, and for the unknot those of Ref.~\cite{Maucher2016}, by testing the bulk untangling dynamics of all of these knots with a wide variety of initial conditions --- in the case of the unknot, a far greater variety than has been used previously. We find that in the bulk a generic unknot, trefoil or figure eight simplifies to a canonical form, but that the wave slapping mechanism at play for large $N$ can cause rates of convergence to vary dramatically. We then characterise the geometry and long-term dynamics of these stationary states and two further examples that we have found, a Whitehead link and a $6_2$ knot, both of which appear to belong to the same `family' as the figure eight knot, sharing with it many dynamical properties. This commonality does not cleave across preexisting knot types, for example torus knots, but rather is a property of the FitzHugh-Nagumo dynamics.

\section{\label{sec:Methodology} Methodology}

\subsection{The FitzHugh-Nagumo model}
The FitzHugh-Nagumo model is given by the pair of nonlinear reaction-diffusion equations

\begin{equation}
\label{eq:FN}
\frac{\partial u}{ \partial t} = \frac{1}{\epsilon}\biggl(u - \frac{1}{3}u^3 -v\biggr) + \nabla^{2} u,\hspace{2em}    \frac{\partial v}{ \partial t} = {\epsilon}(u + \beta -\gamma v) ,
\end{equation}
with $u(\mathbf{x},t)$, $ v(\mathbf{x},t)$ real valued scalar fields. The remaining symbols are model parameters, and here are set to $\epsilon = 0.3$, $\beta=0.7$, $\gamma = 0.5$. These values were originally chosen in Ref.~\cite{Henze1993}, and belong to a parameter regime in which two-dimensional spiral waves rotate rigidly and a simple vortex ring shrinks to a stable finite radius \cite{Courtemanche1990}. As such, they are particularly well suited to the search for stable knots and have been used extensively in the literature \cite{Henze1993,WinfreeChapter,Sutcliffe2003,Maucher2016,Maucher2017,Maucher2018a,Maucher2018b}. With these parameter choices characteristic spatial and time scales are given in arbitrary units (fixed by setting the diffusion constant to one above) by a spiral wavelength, $\lambda_0 = 21.3$, and a rotation period for which we find a value of $T_0=11.14 \pm 0.03$, giving a rotation frequency $f_0$ = 0.0898; this period has been previously been reported as between $T_0 = 11.1$ \cite{Henze1993} and $T_0= 11.2$ \cite{Sutcliffe2003}. One may also define an effective vortex radius $\lambda_0/2\pi \approx 3.4$, a naive estimate of the radius of the stable unknot mentioned above; the actual radius found in Ref.~\cite{Courtemanche1990} is 4.8. Over such a lengthscale one expects short-range inter-vortex repulsion in a generic knotted filament.

\subsection{Simulating bulk FitzHugh-Nagumo dynamics}
\label{subsec:Simulation}
We simulate Eq.~(\ref{eq:FN}) with periodic boundary conditions using the pseudospectral method of Ref.~\cite{Goldstein1996}, in which the linear part of Eq.~(\ref{eq:FN}) is solved exactly in Fourier space via an integrating factor, and nonlinear terms are computed via fast Fourier transform. Thereafter, a fourth order Runge-Kutta timestepping is used. When such a method is employed for diffusive systems, high wavenumbers are damped by an exponential integrating factor, and the system remains numerically stable for time steps beyond those allowed by the Neumann stability criterion \cite{Goldstein1996}. To test the effects of altering gridspacing and timestep we calibrate against the rotation period of a two-dimensional spiral, a quantity known to be sensitive to such choices \cite{Dowle1997}. We have found that for grid spacings of $\Delta x = 0.2,0.4, 0.6$, observing two-dimensional spirals over a time of $T = 5000$ one is able to alter the chosen timestep between $\Delta t = 0.01$ and $\Delta t = 0.14$ with no measurable effect on spiral period. In practice, with the exception of simulations to be discussed in \S~\ref{subsec:Mechanism}, we perform simulations using gridspacing $\Delta x = 0.5$, timestep $\Delta t = 0.1$.

Our use of periodic boundaries complements existing results by removing the effects of no-flux boundary interactions on vortex evolution, allowing us to study long time bulk dynamics without vortex knots breaking at (or nestling into) the boundary. One might worry that although we have removed boundary interactions they have been replaced by the effects of periodic neighbours. Figure~\ref{fig:TypicalSimulation}(a) shows a snapshot of a typical periodic simulation. The structure of the wavefield, shown in orange, is tracked by plotting the level set $u=1.6$. It has a complex topology at lengthscales comparable to that of the vortex filament, shown in red. However further from the filament the shells of wave activity simplify to a series of concentric spheres propagating outward from the location of the filament. This is a consequence of the nonlinear nature of the waves --- when two wavefronts meet they fuse, creating a single cusped front, which is then smoothed by the curvature dependence of front propagation velocity. Provided the simulation domain size remains large in comparison to the dimensions of the vortex filament (and noting that, as we shall see, filament dynamics are typically orders of magnitude slower than wave dynamics), these shells shield the vortex from its periodic neighbours --- the outermost shell passes across the periodic boundary and annihilates itself, leaving the bulk of the simulation untouched. As an example, in figures~\ref{fig:TypicalSimulation}(b, c) we compare snapshots from two simulations of the same knot evolution, in this case the $7_2$ knot, but run in boxes of different sizes. Figure~\ref{fig:TypicalSimulation}(b) is identical to figure~\ref{fig:TypicalSimulation}(a) except that we only show a cross section through the wavefield. Figure~\ref{fig:TypicalSimulation}(c) shows the corresponding knot and cross section taken from the larger simulation. Considering discrepancies between the two wavefields where they overlap (in other words only in the smaller box) we see that differences are localised to a region on the boundary of the smaller box, with the bulk of the two simulations in agreement on this smaller box. As expected, the knot loci themselves are identical. The data shown is taken at time $T=2420$ after initialisation, $O(200)$ vortex rotation periods into the simulation (why we select this particular knot and simulation snapshot for display will be discussed further in \S\ref{sec:UnstableKnots}), during which time (and throughout the remainder of the simulation) the knots from the smaller and larger simulations track one another perfectly. We may thus be confident that our periodic simulation is indeed capturing bulk behaviour. In practice, how large a simulation box one needs to ensure this behaviour will vary depending on knot dynamics and the timescale of simulation. For the results presented here, we find a box size of 174 to be sufficient. 

\subsection{Defining and tracking the vortex filament}
\label{subsec:DefiningTracking}
\begin{figure}[hbtp]
    \includegraphics[width=0.8\columnwidth ]{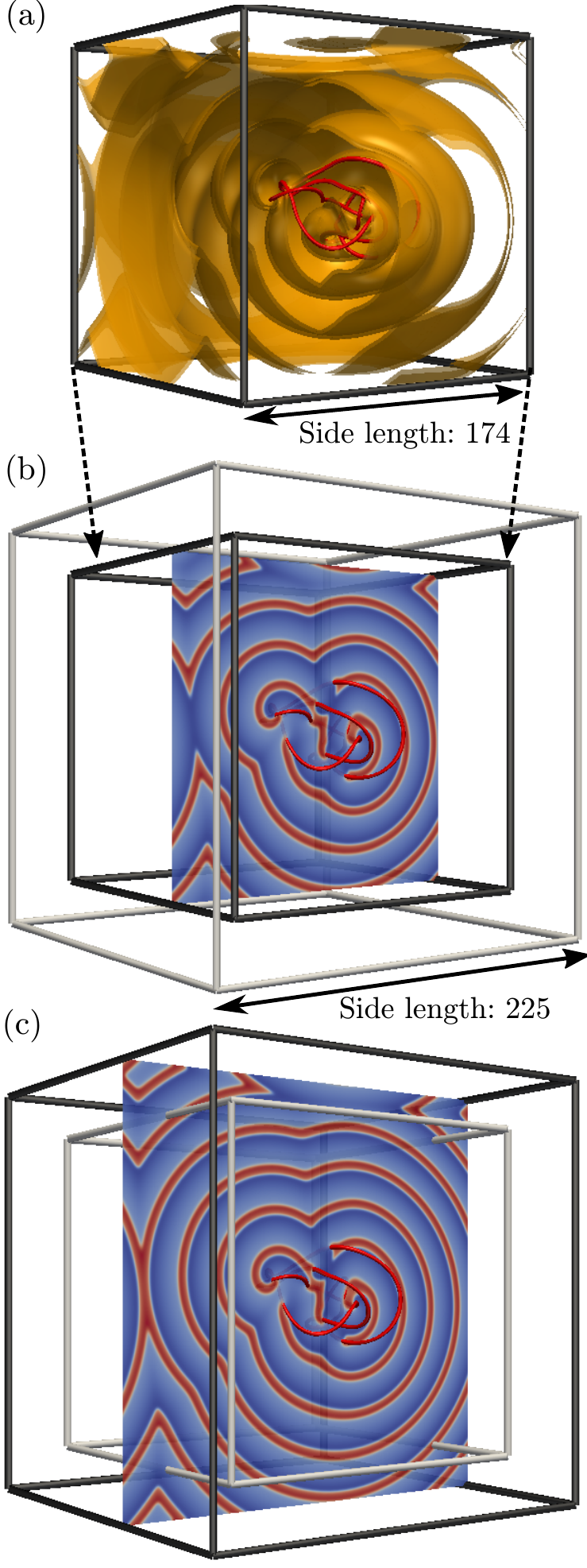}
    \caption{\label{fig:TypicalSimulation} (a) A snapshot of a typical simulation, here of the $7_2$ knot $T=2420$ after initialisation, demonstrating the structure of the wavefield in a periodic simulation box. The level set $|{\bf B}|=0.4$ (red tube) marks the vortex filament, and contours of $ u = 1.6$ mark the location of propagating wavefronts (orange paired surfaces). The near quarter of the wavefronts are clipped to reveal their inner structure. The wavefronts have a complex topology at lengths comparable to that of the vortex, but away from it take the form of simple concentric shells. (b, c) Snapshots of the same $7_2$ knot evolution simulated in boxes of different sizes. (b) replicates (a), with a cross section through the $u$ wavefield shown (wavefronts in red). (c) shows the corresponding knotted vortex and wavefield from the larger box (dark grey simulation box corresponds to cross section shown). Differences in the wavefields are localised to the boundary of the smaller simulation box, ensuring we are indeed capturing bulk dynamics. As expected, the knot loci themselves are identical.}
\end{figure}
Stacking two-dimensional spiral vortices one obtains the simplest example of a vortex filament, one with a straight geometry from which emanates a quasi-two-dimensional `scroll wave' \cite{Winfree1983}. More generally, the vortex filament is a tubular structure with arbitrary geometry, normal cross sections of which resemble spiral vortices whose phase is allowed to vary longitudinally. (In fact even this picture is an idealisation; waves emanating from other sections of the filament can disrupt this local spiral wave structure.) Various operational definitions to extract a one-dimensional curve from this tubular structure have been proposed \cite{Winfree1990,Henze1993,Dowle1997}. Here we follow Refs.~\cite{Sutcliffe2003,Maucher2016,Maucher2017,Maucher2018a,Maucher2018b} and first compute the intermediate quantity
\begin{equation}
\label{eq:ucrossv}
\mathbf{B} = \nabla u \times \nabla v.
\end{equation}
$|\mathbf{B}|$ measures the deviation of $u$ and $v$ contours from colinearity --- it is zero for a planar wave, and only attains substantial nonzero value along the vortex filament. Contours of $|\mathbf{B}|$ thus take the form of tubes. For example, figure~\ref{fig:TypicalSimulation} tracks the vortex filament by showing the level set $|{\bf B}| = 0.4$ in red. To extract a one-dimensional curve from such a tube, we first note that $\bf B$ orients the tube. Stepping along the tube in the direction given by this orientation, we connect maximal values of $|\mathbf{B}|$ in cross sections taken through it, resampling if necessary to give equidistant steps; a similar extraction procedure is detailed in Ref.~\cite{Winfree1990}. This raw curve is then smoothed to remove modes of frequencies comparable to $\lambda$, giving a smoothed curve from which we may compute curvatures and torsions via finite difference. (The choice of lengthscale for filtering is motivated by the observation that the most highly curved stable filament observed, a stable round unknot \cite{Courtemanche1990}, has a circumference comparable to $\lambda$.) Typically we will use the term `filament' to emphasise the one-dimensional curve defined above, and `vortex' when we wish to discuss the full tubular stack of spiral waves surrounding this curve. This definition (and indeed other `instantaneous' definitions \cite{Dowle1997}) gives rise to small amplitude oscillations in the geometry of the filament at period $\sim T_0$ which carry through to derived quantities such as knot length. We shall examine the spectrum of these oscillations in detail in \S\ref{sec:StableKnots}, but in subsequent plots showing length evolution we filter them out for clarity. 

As discussed above, vortex phase varies along the filament, framing our one-dimensional curve, and the twist of this framing may in principle affect both filament motion and vortex rotation period. We track it by computing $\frac{\nabla u}{|\nabla u|}$ along the filament and then smoothing as above. 

\subsection{Initialising a knotted vortex field}

To initialise an arbitrary knotted vortex for simulation we adopt the basic strategy of Refs.~\cite{Sutcliffe2003,Maucher2016} in which a phase field $\phi( {\bf x}) \in \mathbb{S}^1$, ${\bf x} \in \mathbb{R}^3\setminus K$, is constructed which contains a phase singularity with the geometry of some desired vortex knot $K$. Thereafter, the winding of $\phi$ around the specified knotted phase singularity is translated into the winding of $(u,v)$ around the excitation-recovery loop of the FitzHugh-Nagumo model as one encircles the vortex filament via the map $ (u,v) = (2 \cos \phi - 0.4, \sin \phi - 0.4)$.

To construct a phase field $\phi$ containing a singularity along a given curve $K$, we first compute the solid angle function $\omega$ about $K$ using the formula \cite{Binysh2018}
\begin{equation}
    \omega({\bf x}) = \int_{K} \frac{{\bf n}_\infty \times {\bf n} \cdot \mathrm{d}{\bf n}}{1 + {\bf n} \cdot {\bf n}_\infty} \quad \mathrm{mod}\;4\pi,
    \label{eq:SolidAngle}
\end{equation}
where for ${\bf y} \in K$, ${\bf n} := \frac{{\bf y} - {\bf x}}{|{\bf y}-\bf{x}|}$ is the projection of $K$ onto a unit sphere centred on $\bf x$ and ${\bf n}_\infty$ is an arbitrary unit vector. For a discussion of this integral and its numerical properties, including its singular behaviour about points ${\bf x}$ such that ${\bf n} \cdot {\bf n}_\infty = -1$, we refer the reader to Ref.~\cite{Binysh2018}. 

\begin{figure}
\includegraphics{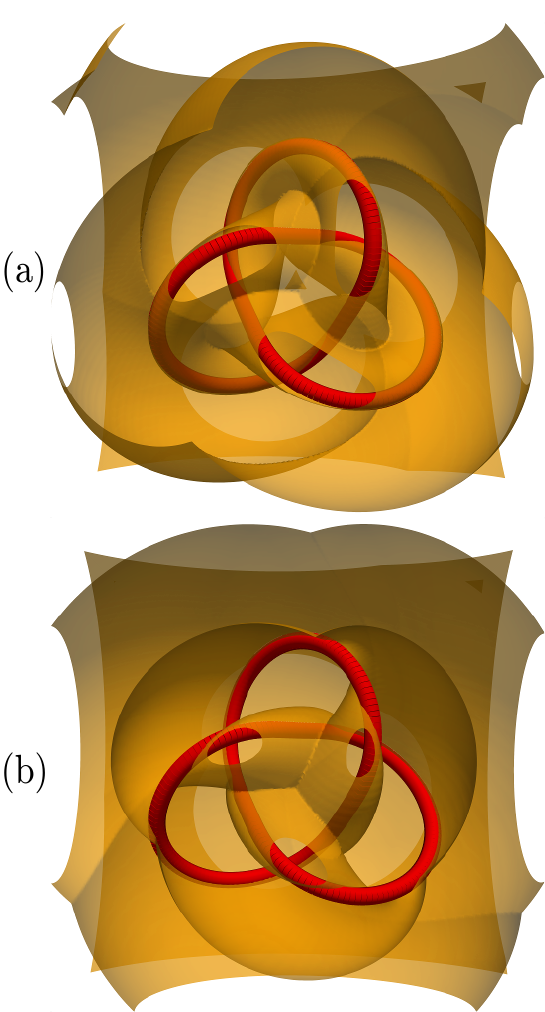}
\caption{Wavefield initialisation about a trefoil knot vortex filament (red curve) using Eq.~(\ref{eq:WavefieldInitialisation}), with the level set $\phi = 0 $ shown in orange. The near half of the level set is clipped to reveal its inner structure. In (a) the definition of $d({\bf x})$ in Eq.~(\ref{eq:WavefieldInitialisation}) is simply minimal distance to the filament. In (b) it is modified by a threefold symmetric sinusoid along the trefoil, effectively adjusting the solid angle framing and local twist rate of the vortex filament.}
\label{fig:WavefieldInitialisation}
\end{figure}

The solid angle contains the necessary phase singularity along $K$, and for the simulations discussed in this paper we use it for initialisation directly by setting $\phi = \omega/2$. We briefly note, however, that the structure of $\omega$ about $K$ does not mirror that of a typical $(u,v)$ wavefield, which consists of a series of approximately equispaced wavefronts radiating outwards from the vortex filament (figure \ref{fig:TypicalSimulation}). Further, this methodology does not give control over the initial twist distribution along the filament, which is set by the intersection of the level set $\omega=0$ with $K$, the `solid angle' framing of $K$ \cite{Binysh2018}. We may control both of these features by modifying $\phi$ as
\begin{equation}
\phi({\bf x}) = k_0 d({\bf x}) + \frac{1}{2} \omega({\bf x}), 
\label{eq:WavefieldInitialisation}
\end{equation}
where $k_0 := 2\pi / \lambda_0$ is the spiral wavenumber and $d({\bf x}):= \mathrm{min}_{{\bf y} \in K} |{\bf y}- {\bf x}|$ is the minimal distance from ${\bf x}$ to $K$. $k_0 d({\bf x})$ increases linearly with distance from the curve, giving a periodic modulation of $\phi$ and hence $(u,v)$ with distance. As an example, figure~\ref{fig:WavefieldInitialisation}(a) shows the wavefield generated using Eq.~(\ref{eq:WavefieldInitialisation}) when $K$ is a trefoil knot. The intersection of the level set $\phi=0$ with $K$, and hence the initial twist distribution, may be controlled by including an offset in the definition of $d({\bf x})$ which varies along $K$. An example of such a modulation, and how it alters the wavefield, is shown in figure~\ref{fig:WavefieldInitialisation}(b). Given, for example, the importance of twist distribution on both rotation frequency and the sproing instability as discussed above (and further explored in this paper), such control is desirable for future work.

Initialisation geometries for the knots considered here were constructed from those found in \emph{KnotPlot} \cite{KnotPlot}, which in turn are based on Rolfsen's knot table. In the absence of existing work on high crossing number knots in the FitzHugh-Nagumo model, there is no compelling reason to choose one set of initialisation geometries over another; for example, an alternative choice would be to use configurations of ideal ropelength \cite{Kleckner2016,Maucher2017,Cantarella2011}. We use Rolfsen's configurations as, with the exception of the torus knots (whose evolutions we will compare to existing results \cite{Maucher2017}) the geometries do not possess any symmetries. If strands of the knot are initialised closer to one another than the vortex radius $\lambda_0 /2\pi$, reconnections may occur in the first $\Delta T \approx T_0$ of simulation, before the wavefield about the knot is established \cite{Maucher2016}. To ensure this does not occur, given an initialisation geometry $K$ we scale isotropically such that the longest side of $K$'s bounding box occupies $80\%$ of the simulation box size. As the simulation box is $O(50)$ times the size of the vortex radius, this ensures reconnections do not occur during initialisation.

\section{\label{sec:UnstableKnots}Unstable Knots}

\begin{figure}
    \includegraphics[width=0.9\columnwidth]{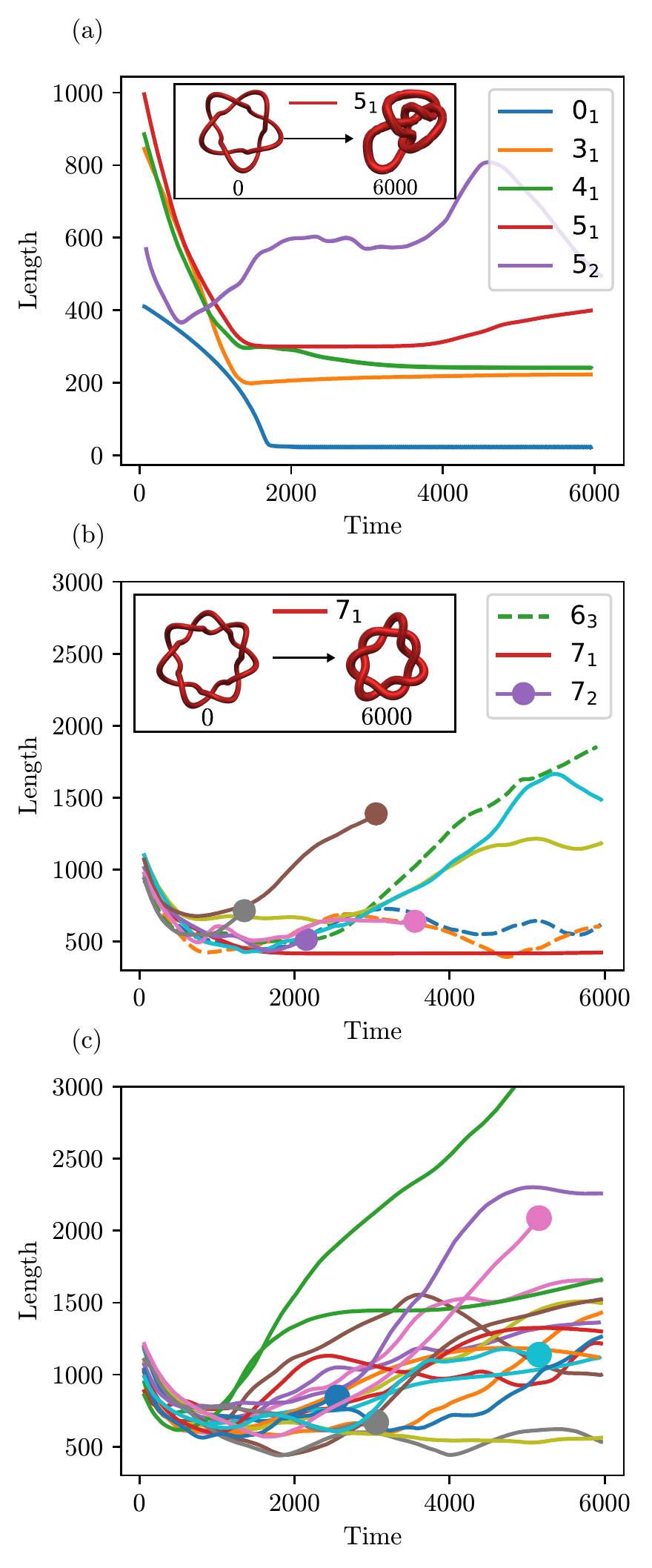}
    \caption{ Length evolution of knotted vortices up to and including crossing number $N=8$, with reconnection events indicated by curves which terminate early with a circular marker. Knots included in the legend are further discussed in the text. Note the difference in scale between the first and subsequent panels. (a) $N\leq5$. (b) $N=6$ (dotted lines) and $N=7$ (solid lines). (c) $N=8$. The unknot ($0_1$), trefoil ($3_1$) and figure eight ($4_1$) settle to a stable length and fixed geometry (see section \ref{sec:StableKnots}). However, beyond this, generic behaviour is an initial period of contraction, followed by length increase over longer timescales. Insets show the geometries resulting from the destabilisation of the initially symmetric $5_1$ and $7_1$ torus knots.} 
\label{fig:UnstableKnotsLength}
\end{figure}

\begin{figure}
    \includegraphics[width=\columnwidth]{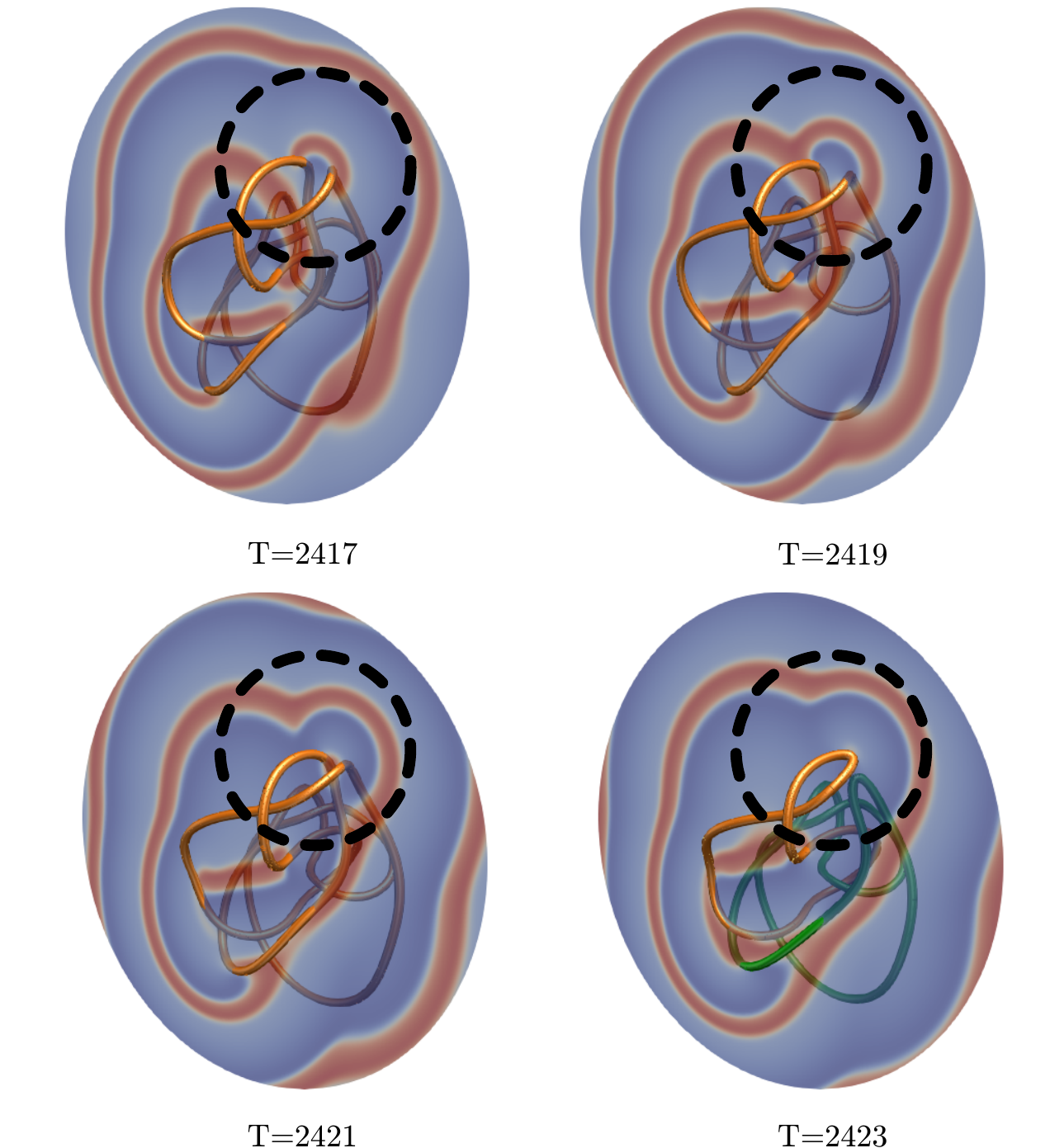}
    \caption{Reconnection of the $7_2$ knot at $T=2423$ (orange curve, then orange and green curves after reconnection event) shown in $\Delta T = 2$ increments. A pair of anti-parallel strands (circled in black) interact, generating a wavefield locally similar to that of the stable unknot; shown is the value of $u$ in a cross section through the knot, with high $u$ value coloured red. The wavefield from the remainder of the knotted filament, shown entering the circled region at $T=2417$, impinges upon these strands causing them to destabilise and reconnect. Note that the wavefield at $T=2420$ is also shown in figure~\ref{fig:TypicalSimulation}.}
\label{fig:Reconnections}
\end{figure}

Fascinating recent numerical experiments on the evolution of unknots initialised in complex geometries showed that the FitzHugh-Nagumo dynamics is capable of simplifying an initially tangled unknot to a unique circular curve, without strand crossings~\cite{Maucher2016}. These intriguing examples, coupled with two further instances of simplification in the case of the trefoil \cite{Sutcliffe2003}, as well as some indirect evidence of the same behaviour for the Hopf link \cite{Maucher2018b} (and a series of preliminary results on various links in Ref.~\cite{Henze1993}), naturally invite the speculation that such simplification is generic to any knot. To investigate the dynamics of a generic knotted filament, and establish whether this is indeed the case, in figure~\ref{fig:UnstableKnotsLength} we show a survey of the length evolution of all knots up to and including crossing number $N=8$, with particular curves that we discuss further highlighted in the legend. Excluding chiral variants, there are thirty-six such knots. 

Initial behaviour across all knots is contraction; this behaviour is purely a result of curve geometry, reflecting an effective positive line tension for filaments well separated from any interactions \cite{Biktashev1994}. Our curves are initialised with their strands separated by several vortex radii. Over the first few vortex rotation periods the wavefield establishes itself around the vortex filament (over such a timescale the filament may be considered stationary) with the resulting collision interface disjoint from the filament itself. Thereafter each segment of the filament initially moves effectively in isolation from its neighbours. 

Over longer timescales, however, we find that this initial contraction does not generically lead filaments to settle to a canonical form or a fixed length, and further that their topology is not always preserved. We observe reconnection events in eight of the thirty-six cases, indicated by those curves terminating in circular markers in figure~\ref{fig:UnstableKnotsLength}. The first of these occurs at $N=7$, with four of the seven $N=7$ knots and four of the twenty-one $N=8$ knots exhibiting reconnections.

Figure~\ref{fig:Reconnections} shows the reconnection event in the $7_2$ knot occurring at $T=2423$ in $\Delta T=2$ increments, with the region where the reconnection occurs circled in black. A pair of neighbouring anti-parallel segments are directly impacted by waves emanating from the rest of the knotted filament, causing them to destabilise and reconnect --- the same wave slapping mechanism responsible for the annihilation of a small unknot by a larger one observed in Ref.~\cite{Maucher2018a}. In cross section, the wavefield generated by the anti-parallel segments (that section of the wavefield ending at the circled anti-parallel segments in the $T=2417$ panel of figure 4) locally resembles that of a stable unknot, or indeed of a pair of oppositely signed two-dimensional vortices \cite{Courtemanche1990}. However this stable structure is additionally impinged upon by the wavefield of the rest of the knot (shown entering the circled region at $T=2417$). It has previously been noted that the rotation period of a stable unknot is 14\% greater than $T_0$ \cite{Maucher2018a}, and as such the stable unknot is vulnerable to wave-slapping induced annihilation, as it cannot `fend off' a wavetrain of period $T_0$. This same argument applies to the anti-parallel segments discussed here, but the resulting topological change is reconnection. In addition to the similarity of the wavefields between the coaxial unknots of Ref.~\cite{Maucher2018a} and the situation discussed here, the relative filament motion is also the same; in both cases, the perturbed filaments are drifting away from the impinging wavefield when topological change occurs. This geometric detail is important, as the velocity of a stable unknot ($0.3$ \cite{Maucher2018a}) is a substantial fraction of the wavespeed in the medium ($1.9$) and as such Doppler shift may compensate for reduced unknot frequency. For the geometry here, however, the two effects can only compound one another. 

Topology changes previously reported in the literature have primarily occurred at simulation boundaries \cite{Maucher2017, Maucher2018b}, and one might be concerned that this reconnection is also an artefact of a finite simulation box. The snapshot of a $7_2$ simulation at $T=2420$ shown in figure~\ref{fig:TypicalSimulation} was taken from the same data shown in figures~\ref{fig:UnstableKnotsLength}, \ref{fig:Reconnections}, and was chosen specifically to emphasise that this is not the case. Enlarging the simulation box to side length $225$ as shown in figure~\ref{fig:TypicalSimulation} yields identical knot evolution, including the reconnection event.

Although we have focused the above discussion on the reconnection in the $7_2$ knot, analogous findings hold for the other cases. A detailed understanding of the wavefield evolution leading to such reconnections, and why they are seen here only for $N>5$, is lacking (we shall see examples of similar wavefield induced effects for small $N$ in \S \ref{sec:StableKnots}), but as $N$ increases one generically expects a more complex wavefield surrounding the knot as strands become closely packed over one another. As such, we lose the ability to picture segments of the knot as isolated, or even as interacting solely with a unique `nearest neighbour' region, but instead must think of them analogously to the vortex ring buffeted by external waves.

Even in the absence of reconnections, we only see stable states for the unknot, trefoil and figure eight knots; we shall discuss the robustness and detailed geometry of these states in \S \ref{sec:StableKnots}. For $N>4$ knots do not simplify. After the initial relatively rapid contraction, we generically see expansion over a much longer timescale of order several hundred vortex rotation periods followed by periods of irregular evolution including possible further expansion. Although the broad trend is to faster expansion at higher crossing number, the variability of behaviour between knots of the same crossing number suggests that initialisation geometry is just as important in determining long time evolution. In particular, we note the discrepancy between the behaviour of a typical knot with no initial symmetry and a torus knot initialised with high symmetry. Although a typical knot shows length increase by time $T\approx2000$, for the $5_1$ knot shown in figure~\ref{fig:UnstableKnotsLength} this increase is only visible by $T\approx4000$, and is not evident for the $7_1$ even by $T\approx6000$, although the inset curve geometries demonstrate the knot has indeed destabilised (that such destabilisation occurs, but only after long time simulations, clearly demonstrates the necessity of simulating for many hundreds of rotation periods before drawing conclusions about the dynamics of this system). In the next section, we shall investigate the mechanisms of both the dramatic length increases seen in a generic knot, and of the deviation of the torus knots away from initialisations of high symmetry. 
\begin{figure*}
    \includegraphics[width=0.85\textwidth]{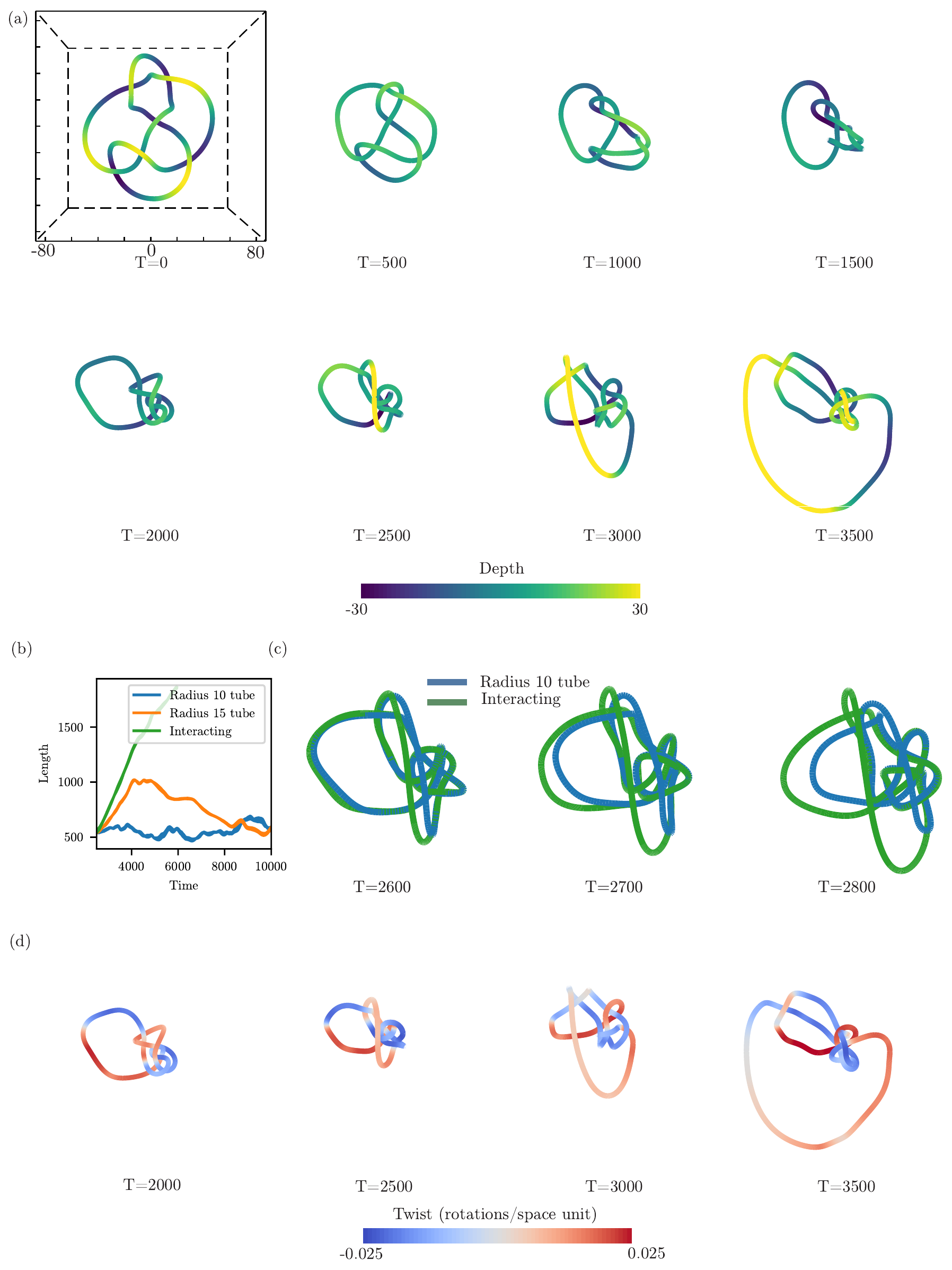}
\end{figure*}
\subsection{The mechanism of vortex knot length increase}
\label{subsec:Mechanism}
\begin{figure*}[hbtp]
    \caption{ Dynamics of the $6_3$ knot. (a) From $T=0$ to $T=2000$ the knot contracts and flattens, behaviour caused by the intrinsic curvature driven dynamics of an isolated filament mimicking a line tension. From $T=2000$ onwards length increases, with a single arm of the knot rapidly expanding outwards from an otherwise tightly packed core. (b) Comparison of the length evolution of the fully interacting $6_3$ vortex filament at $T=2000$ (green) to a copy encased in a `glass tube' of radius 10 (blue) or 15 (orange). With long-ranged interactions removed, the knot does not expand, but rather settles to a length of $\sim 400$ -- $600$. (c) Initial divergence in geometries between the interacting $6_3$ knot (green curve) and the radius $10$ tubed one (blue curve). Divergence does not occur globally, but is localised to two distinct expanding segments outside the lengthscale defined by the tube. (d) Distribution of filament twist during knot expansion. The expanding arm of the knot has twist values well below the $0.024$ rotations per space unit threshold for the sproing instability, and is less highly twisted than other, non-expanding, segments of the knot, ruling out this instability as the cause of length increase.}
\label{fig:63evolution} 
\end{figure*}
Exploring the knot geometries corresponding to the generic length increases seen in figure~\ref{fig:UnstableKnotsLength} one finds that, despite the variety of behaviour across knots, the increase occurs via a common mechanism in which isolated strands of the knot rapidly expand outwards from a tightly packed core region forming the rest of the knot. We illustrate this behaviour for the example of the $6_3$ knot in figure~\ref{fig:63evolution}(a). The same wave slapping mechanism driving reconnection events has also been proposed as a non-local mechanism for persistent knot length increase; in this context when the collision interface intersects a section of the knot, wave-vortex interactions drive that section outwards \cite{WinfreeChapter,Sutcliffe2003}. The interaction does not have an intrinsic lengthscale, as waves in the medium do not decay. Instead, its range depends upon the geometry of the knot and the accompanying collision interface. For the $6_3$ knot shown in figure~\ref{fig:63evolution} one may verify that this surface intersects the expanding arm, suggesting that wave slapping may be at play. 

Given the potential importance of this mechanism, we would like to establish that it is really driving knot expansion, rather than simply being correlated with it. To do so, we investigate the effects of abruptly removing long-ranged interactions entirely, by numerically encasing the filament in a `glass tube' of moderate radius which moves with the filament and fuses when two knot segments approach one another \cite{Winfree1983b}. With this construction short-range inter-filament repulsion, and geometry (including twist) mediated filament motion are preserved but long-ranged interactions are cut out. Using it, we may compare the evolution of a filament both with and without long-range interactions. A suitable radius for the tube is suggested by previous estimates of vortex radius in the literature, as well as the naive estimate $\lambda_0/2 \pi \approx 3.4$: Ref.~\cite{Courtemanche1990} directly measures a stable vortex ring radius of $4.8$, suggesting a vortex radius of $\sim~5$, and Ref.~\cite{Maucher2017} estimates a radius of $5.9$ by matching ideal ropelength \cite{Cantarella2011} and measured trefoil lengths. To implement this construction numerically we simulate only within a tube of lattice points about the filament (the filament itself being constructed as in \S\ref{sec:Methodology} C). In principle the details of the boundary conditions between the vortex and tube must be considered, however we have found that provided we use a tube radius above the vortex size estimates above, such details do not alter the geometry driven motion of the vortex, a reflection of its localised nature. This observation allows us to sidestep a sophisticated finite-element scheme (the spectral method discussed in \S~\ref{subsec:Simulation} only being valid for a periodic box), and instead simply use a finite difference method, with $(u,v)$ values for points outside of the tube set to their fixed point values $(-1.03,-0.66)$. The tube must move with the vortex, however we note that this need only happen on the (slow) timescale of vortex motion; we may allow motion in a fixed tube for a time $O(T_0)$, after which the tube is recentred around the vortex. The points brought into the tube at its boundary during this procedure are again set to fixed point $(u,v)$ values. In practice we typically use a conservative tube radius of $\sim 10$ with gridspacing $\Delta x = 0.5$ and timestep $\Delta t = 0.01$, with a finite difference scheme in which the Laplacian is computed using a seven point stencil and both reaction and diffusion terms are evolved using fourth order Runge-Kutta timestepping. The timestep above is chosen as it gives results identical to those of the spectral method using $\Delta t = 0.1$ when measuring the two-dimensional spiral vortex period. 

Figure~\ref{fig:63evolution}(b) contrasts the length evolution of the fully interacting $6_3$ knot with a copy of it encased in the tube, with initial conditions for both taken at $T=2500$, midway through knot expansion. Upon removing long-ranged interactions we no longer see a dramatic increase in knot length. Instead, the length of the tubed knot stabilises at $\sim 400- 600$. The details of this stabilisation vary depending on the radius of the tube, but the final lengths obtained are approximately the same across radii. Using the core size estimate of Ref.~\cite{Maucher2017}, the ideal ropelength of the $6_3$ knot is $340$~\cite{Cantarella2011}, and thus the tubed knot is relatively tightly packed. Over longer timescales ($\Delta T=8000$ shown for the radius $10$ tube of figure~\ref{fig:63evolution}(b)) the tubed $6_3$ does not reach a fixed geometry, but rather undergoes a compact tumbling motion, as the binormal component of filament motion causes segments of the knot to work over one another, though without further substantial length change. In figure~\ref{fig:63evolution}(c) we explore the initial divergence in geometry between the fully interacting and tubed knots. We see that it does not occur globally, but is localised to distinct expanding segments of the interacting $6_3$ which lie separate to the knot core region and are responsible for global length increase; these same segments are those which intersect the collision interface. Within the core region, segments of the filament are packed closer than the spatial cutoff we have defined, and there is no immediate divergence between the interacting and tubed knots. By contrast, removing long-range interactions allows distant segments of the tubed knot to evolve under their intrinsic dynamics, unaffected by wave-vortex interactions, and so shrink towards the core region. Thus a wave slapping mechanism accounts for global changes in knot length and also for the geometry of where they occur. 

In local geometric models of filament motion a mechanism by which filament length may stabilise or increase, despite an effective positive line tension, is via the `sproing' instability \cite{WinfreeChapter} in which, above some critical local twist threshold, an initially straight filament expands into a helix --- for the FitzHugh-Nagumo model with the parameter values used here, Ref.~\cite{Henze1993} reports this threshold at 0.024 rotations per space unit for a straight filament. This instability has been proposed to account for the halting of links at lengths greater than hard core repulsion on the scale of the vortex radius would suggest \cite{WinfreeChapter}, and for the destabilisation of symmetric torus knots \cite{Maucher2017}. We may rule out sproing as a driver of the dramatic length increases seen in generic knots by noting that, as a local geometric mechanism, its effects were present in the tubed knot discussed above. For further confirmation, we may also examine the twist distribution along the fully interacting filament during knot expansion. Figure~\ref{fig:63evolution}(d) shows this distribution for the $6_3$ knot; we see that the expanding arm of the filament consistently has twist values well below the sproing threshold and, further, that other sections of the knot are more highly twisted, yet do not show the same length increase. In fact, twist values along the entirety of the knot are consistently below the sproing threshold, an observation also made for the early short time simulations of Refs.~\cite{Henze1993,WinfreeChapter}. 

\begin{figure}
    \includegraphics[width=\columnwidth]{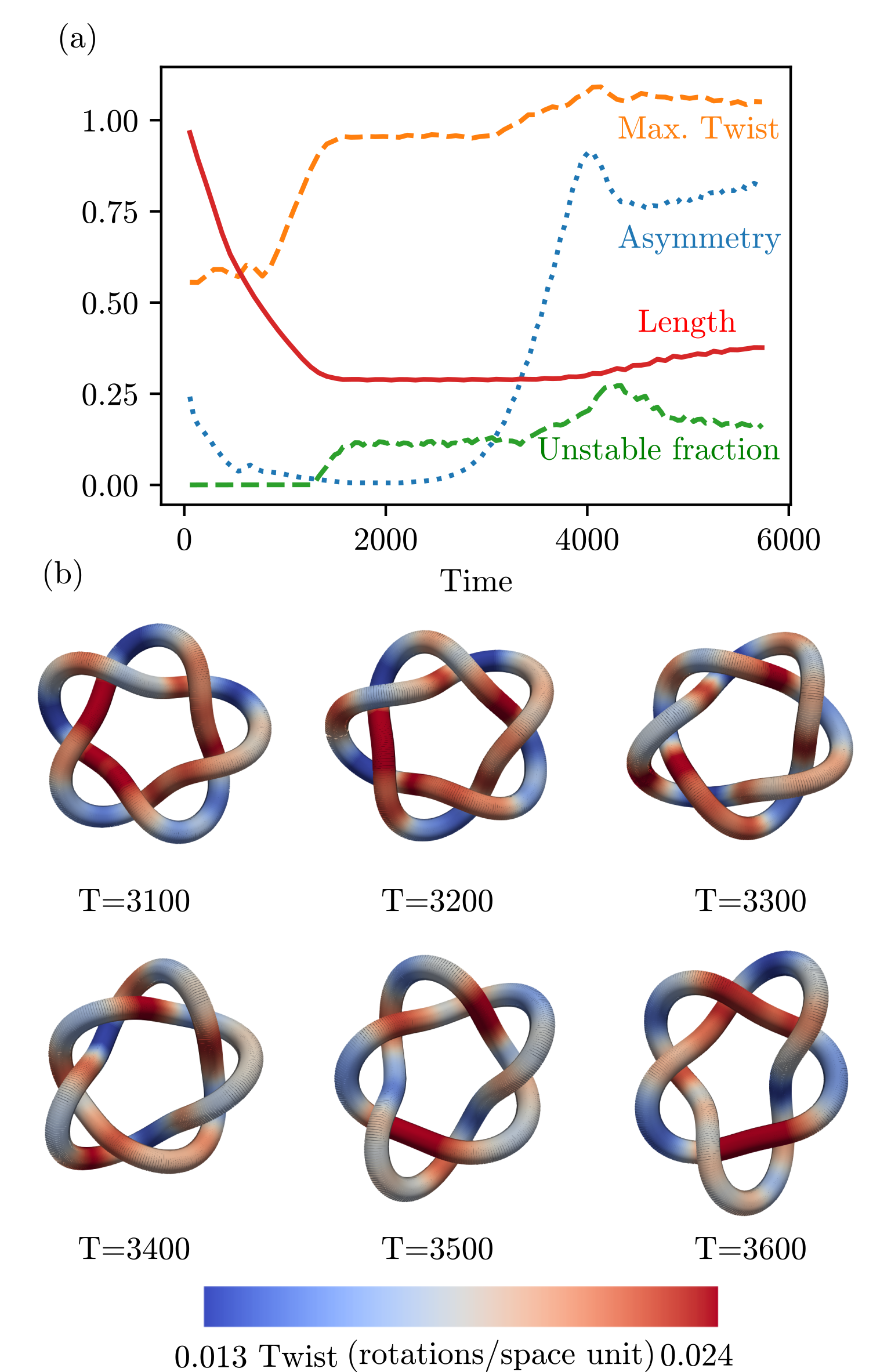}
    \caption{ The role of the sproing instability in the destabilisation of the $5_1$ torus knot. (a) shows the (normalised) length evolution of the $5_1$, alongside a measure of its asymmetry. Shown also is the maximum absolute twist along the knot as a fraction of the 0.024 rotations per space unit sproing threshold, and the fraction of arclength which attains 90\% of this threshold. The twist threshold is reached as knot length plateaus, but no sproinging instability is observed; instead the knot gradually destabilises over several hundred rotation periods. (b) shows the geometry of the knot destabilisation, coloured by twist. Rather than a helical instability developing in regions of high twist, the whole knot transitions to a twofold symmetric form.}
\label{fig:TorusDestabilisation}
\end{figure}

Although not a driver of generic knot length increase, this last observation suggests that the sproing threshold may still have dynamical importance as a stabiliser against curvature induced length decrease, or play a role in the destabilisation of symmetric torus knots. In figure \ref{fig:TorusDestabilisation} we study the destabilisation of the $5_1$ torus knot, originally presented in figure \ref{fig:UnstableKnotsLength}, in more detail. Figure \ref{fig:TorusDestabilisation}(a) shows the evolution of a measure of the asymmetry of the knot, defined by taking the power spectrum of the knot's curvature as a function of arclength, and computing the fraction of the power in modes which do not respect the underlying symmetry (fivefold in this case). Alongside it we show the evolution of both the maximal twist, expressed as a fraction of the $0.024$ rotations per space unit sproing threshold discussed above, and the fraction of the arclength of the $5_1$ which attains a twist greater than $90\%$ of this threshold. We first note that the order of events is broadly consistent with sproing threshold playing a role in the dynamics. After an initial period in which the knot flattens and the twist remains roughly constant, maximal twist increases until it attains the sproing threshold, thereafter remaining constant; this threshold is attained as the length of the $5_1$ stabilises. However, it is several hundred rotation periods before we see the subsequent loss of symmetry. This timescale suggests that it is not the case that the knot hits the sproing threshold, and then destabilises; Ref.~\cite{Henze1993} notes that the timescale for sproinging to occur is typically only a few rotation periods. Furthermore the geometry of the destabilisation is inconsistent with sproing instability. In figure~\ref{fig:TorusDestabilisation}(b) we show the knot as it destabilises, coloured by twist. We fail to see helical sproinging along the highly twisted segments of the knot; instead the whole form collapses to a twofold symmetric shape. A similar deformation is seen in the $7_1$ (see inset of figure \ref{fig:UnstableKnotsLength}) and has been noted in early simulations of initially symmetric triply-linked rings \cite{Henze1993}, where its cause was attributed to an interplay between the sproing threshold and inter-filament interactions. Overall, then, it appears the sproing threshold acts to halt knot shrinkage, but that subsequent destabilisation cannot be directly attributed to the sproing instability. 

\section{\label{sec:StableKnots}Stable Knots}
\begin{figure*}
    \includegraphics{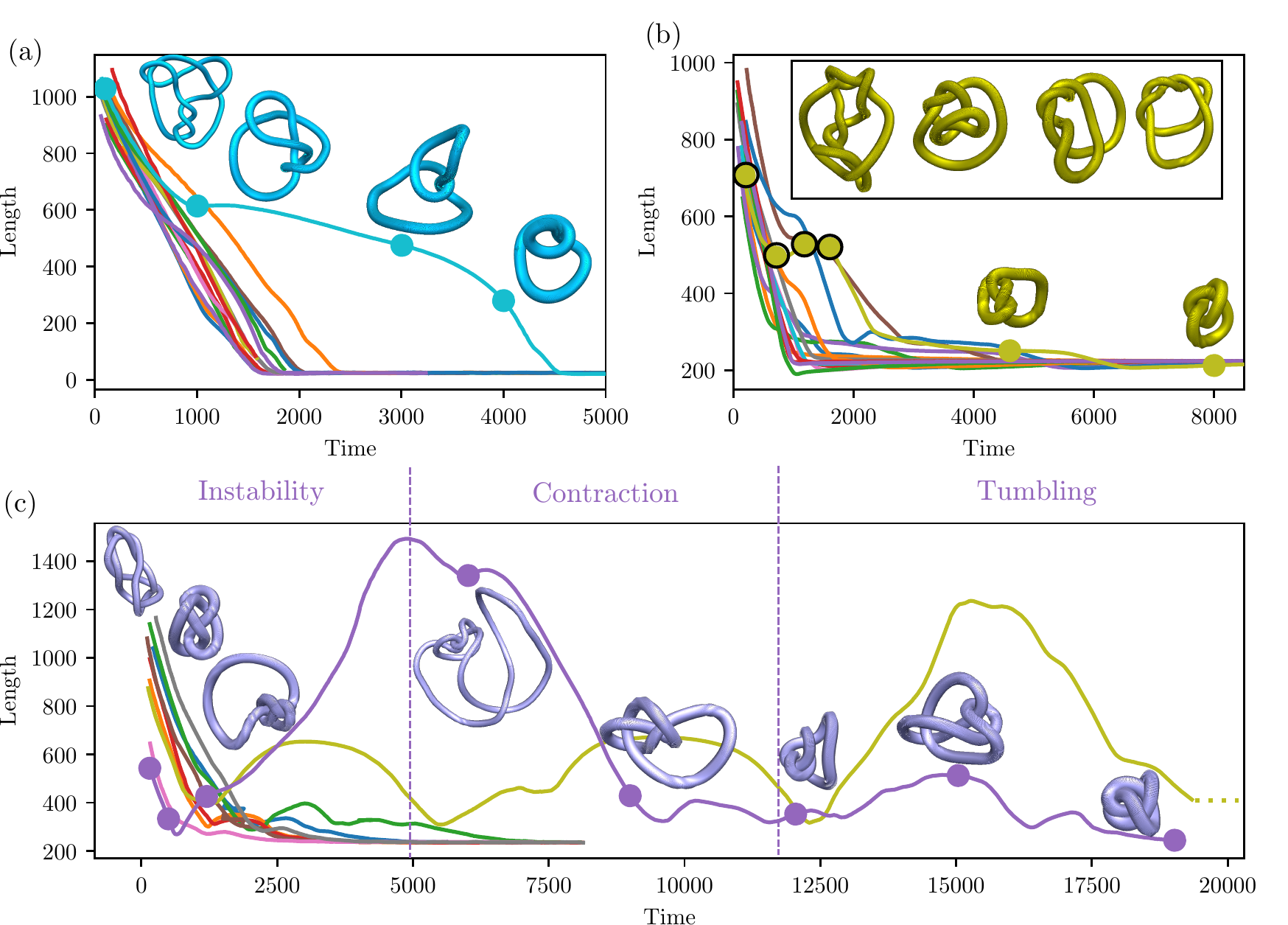}
    \caption{ Untangling dynamics of the (a) 18 unknots, (b) 17 trefoils and (c) 9 figure eight knots formed by performing single strand crossings on the higher crossing number knot geometries of \S \ref{sec:UnstableKnots}. (a) All unknots simplify to a unique round geometry without reconnection events. Length decrease is monotonic, however there is some variation; the geometry of one particularly slow decay is shown in the inset, displayed at times indicated by the solid markers. (b) All trefoil geometries simplify to a unique stable state, however there is greater variation across decays than for the unknots, with periods where knot length actively increases (boxed inset, circled markers). (c) Of the $9$ tangled figure eights simulated, $7$ settle rapidly to a stable state. However, over $T=20000$ one example fails to converge and another converges only after going through prolonged periods of length increase, contraction and irregular `tumbling' dynamics.} 
\label{fig:SecretFourOneLength}
\end{figure*}

In \S \ref{sec:UnstableKnots} we showed that the speculation that a generic knotted vortex might simplify to a canonical form --- a speculation previously evidenced by promising `untangling' results for the unknot \cite{Maucher2016} and a few further examples of simplification in low crossing number knots and links \cite{Sutcliffe2003,Maucher2018b} --- is not borne out for $N>4$. In the search for stable knots, recent numerical experiments found that knots and links could be stabilised through proximity to a no-flux boundary~\cite{Sutcliffe2003,Maucher2017}. Primarily the examples shown were for torus knots and links, although the figure eight knot and Borromean rings were also briefly given as non-torus examples. In contrast to the untangling of unknots, these boundary stabilised states of more complex vortices were not established to be `basins of attraction' for generic initial geometries, but rather were obtained from highly symmetric initial vortex line geometries. In \S \ref{sec:UnstableKnots} we also saw that in the case of torus knots more complex than the trefoil such states are not stable in our bulk simulations. By contrast, for the stability of the trefoil and figure eight knots we now show that a much stronger statement than has been made previously is true: in the bulk, a generic trefoil or figure eight simplifies to a canonical form, analogously to the unknot. The states are the same as those found in the survey of figure \ref{fig:UnstableKnotsLength} and also appear to be the same as those found near a reflecting boundary. In addition, we strengthen the results of Ref.~\cite{Maucher2016} and demonstrate them to be independent of a no-flux boundary by testing the bulk untangling dynamics of the unknot with a far greater variety of initial conditions than has been used previously.

All knots may be converted into the unknot by performing strand crossings. The minimal number of strand crossings needed to convert a knot into the unknot is called its unknotting number. Of the knots with $N\leq8$ there are $18$ with unknotting number $1$; that is, they can be converted to the unknot by a single strand crossing. By analogous single strand crossings one can also target the trefoil or figure eight knots: for $N\leq8$ there are $17$ that convert to the trefoil and $9$ to the figure eight under a single strand crossing. Beginning with the knot geometries of \S \ref{sec:UnstableKnots}, we use these crossings to provide an assortment of initial tangled geometries for unknots, trefoils, and figure eights, and study their evolution. Figure~\ref{fig:SecretFourOneLength}(a) summarises the results of these simulations for the tangled unknots. We find in all cases that the initially tangled vortex transforms to a unique stable ring and that the dynamics does not involve any reconnections. The typical dynamics is an approximately constant rate of length contraction, although this is not rigorous and there is some variation. In particular, in one example (obtained from the $8_{11}$ knot) there is a substantial period of pause where length decreases much more slowly than is seen on average; snapshots of the geometric evolution of this curve are shown as insets. 

Figures~\ref{fig:SecretFourOneLength}(b) and (c) show results for the trefoil and figure eight knots. As with the unknot, for the trefoils we see simplification without reconnection to a unique steady state, although there are perhaps more examples showing periods where the length is not decreasing; one such is illustrated by the inset figures. However, for the figure eights the dynamics is rather more complicated; $7$ of the $9$ initialisations rapidly converge to a unique stable state, but $2$ show prolonged periods of length increase as well as of contraction, with one of them failing to converge over the times simulated. In the example highlighted in figure~\ref{fig:SecretFourOneLength}(c), we see that the initial period of expansion is due to a single arm of the knot rapidly expanding outwards from an otherwise tightly packed core, caused by the same wave slapping as described in \S\ref{sec:UnstableKnots}. This expansion continues for many hundreds of rotation periods and results in a total increase of several times the initial knot length. In addition, the subsequent period of contraction does not lead directly to a stable shape, but rather produces an extended period of `tumbling' dynamics in which the length fluctuates erratically before eventually settling to the final steady state. The total time that this dynamics plays out over greatly exceeds that of the typical unknot. 

These results bridge the gap between the simplification of the unknot discussed in Ref.~\cite{Maucher2016} and our own findings for high crossing number knots by showing that, although clearly neither the untangling dynamics nor the geometries giving rise to wave slapping instability are fully understood, the same mechanisms dominating high crossing number knot behaviour also play an important role in determining low crossing number behaviour; wave slapping can totally disrupt the appealing picture of a dynamics which monotonically decreases knot length even when a stable target state exists. The results also demonstrate the importance of initial conditions on long-term knot evolution; even given the existence of a stable state, the difference between a `good' and `bad' initial starting state may lead to an order of magnitude difference in the time taken to reach that stable state. 

Another notable example of the importance of initial conditions comes from the observation that the boundary stabilised trefoil knot actually exists in two distinct stable configurations~\cite{Maucher2017}. The first, which we denote the $3_{1,1}$, has the geometry that the tangled trefoils of figure \ref{fig:SecretFourOneLength} evolve to. The second, which we denote the $3_{1,2}$, is not reached by our tangled trefoils. This state was constructed in Ref.~\cite{Maucher2017} from an exactly twofold symmetric initial vortex filament, and preserves this symmetry in the final reported state. Although, as we have seen with torus knots, highly symmetric boundary stabilised states may not exist in the bulk, in our own simulations we have found the $3_{1,2}$ to be accessible in the bulk using an initial configuration with only approximate twofold symmetry, and have confirmed its stability up to $T=12000$. Thus, although this twofold symmetric $3_{1,2}$ indeed appears stable, the results of our tangled trefoil simulations suggest that it has a small basin of attraction. Taken together, the above results suggest that, although we have seen that the stability of higher crossing number knots is not the norm, stable geometries may nevertheless exist in the bulk, but that when hunting for them we should not use any carelessly chosen initial configuration, but ought to be more selective in which initial geometries we use. For hints as to what those geometries might be, we now investigate in detail the properties of the stable knots we have found thus far.

\subsection{Properties of stable knots}
\begin{figure*}[hbtp]
    \includegraphics[width=0.8\textwidth]{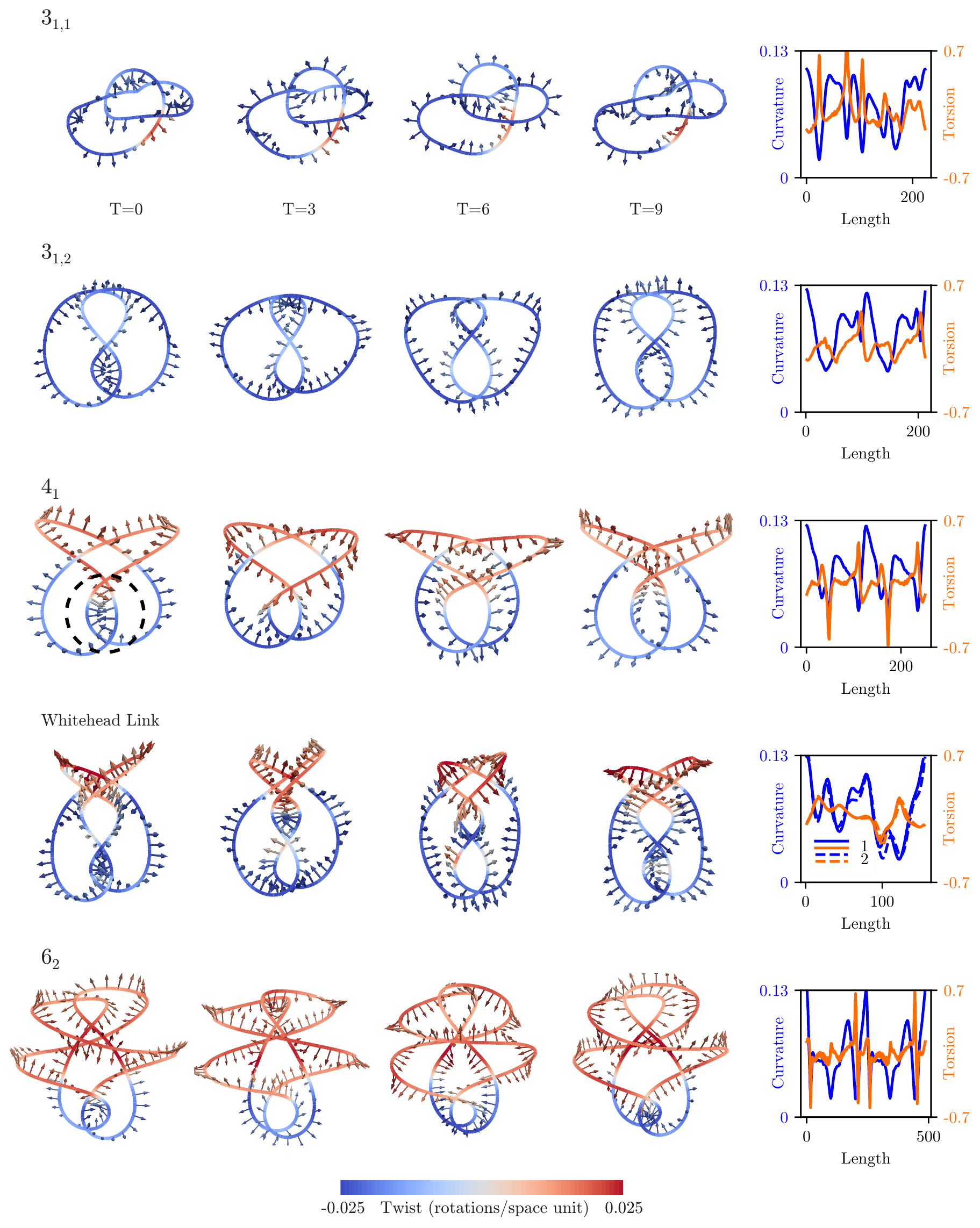}
    \caption{Geometries, vortex framings and twist distributions of our stable knots. Vector fields along curves indicate vortex framings, and are shown at four successive times across a (approximate) vortex rotation period. Curvatures and torsions shown correspond to the $T=0$ panels (there is slight intra-period variation) with the zero of arclength fixed to maximal curvature values. With the exception of the $4_1$ knot for which there is no distinction, all knots shown are the `right handed' chiral variant --- they rotate in a right handed sense about their direction of motion (down the page). Dotted circle highlights a half turn of a helix in the figure eight geometry, which may be extended to several half turns to give the initialisation geometries for the stable Whitehead link and $6_2$ knot shown. }
\label{fig:StableKnots}
\end{figure*}

Figure~\ref{fig:StableKnots} shows the geometries, curvature and torsions as a function of arclength, vortex framings and twist distributions of the stable $3_{1,1}$, $3_{1,2}$ and figure eight knots. The evolution of their vortex framings at four successive intervals over a (approximate) vortex rotation period are indicated by the vector fields along the curves. Curvatures and torsions shown correspond to the geometries in the far left, $T=0$ panels (as discussed in \S \ref{subsec:DefiningTracking} there is slight intra-period oscillation), with the arbitrary zero of arclength fixed to coincide with maximal curvature values. We first note the striking twofold symmetry of both the $3_{1,2}$ and the $4_1$ knots; this symmetry is not a remnant of initial conditions, but emerges from the underlying dynamics. By contrast, the $3_{1,1}$ lacks any threefold symmetry. This is especially notable given that this state was reached starting from an exactly threefold symmetric torus knot geometry in \S \ref{sec:UnstableKnots}. It appears that the boundary stabilised trefoil reported in Ref.~\cite{Maucher2017} also lacks threefold symmetry, although it is unclear why this loss of symmetry does not occur for boundary stabilised torus knots of higher crossing number. A second striking feature of these stable knots is the tight synchronisation of the evolution of their framings. The framings of closely separated segments of the filament mesh \cite{Henze1993}, the wavetip emanating from one segment being consistently met by a wavetip emanating from a spatially neighbouring segment, resulting in travelling waves of tightly synchronised wave activity running the length of the knot in a periodic fashion. The pattern is evident in the $3_{1,2}$ and figure eight knots, but is also present in the $3_{1,1}$, most clearly when one focuses on one of its three relatively straight segments; the framing of the curved lobes is twisted such that it meets the rotation of the wavetip emanating from the straight segment. Again, this meshing is an emergent property of the stable knot. 
\begin{figure}[tbp]
    \includegraphics[width=0.9\columnwidth]{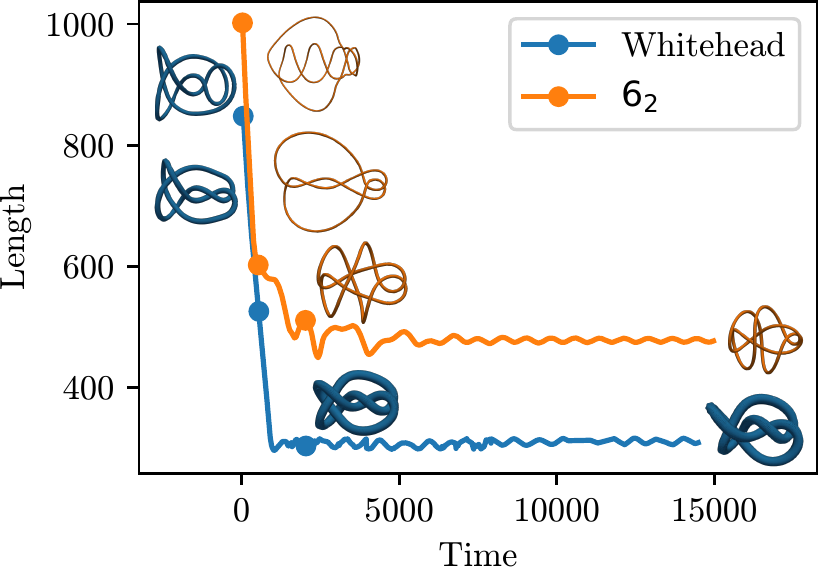}
\caption{Length evolution of the Whitehead link and the $6_2$ knot with initialisation geometries made by extending the structure of the stable $4_1$ knot. Insets correspond to marked times.}
\label{fig:Whitehead_6_2}
\end{figure}

The similarity of the geometries and vortex framings of the $3_{1,2}$ and figure eight is suggestive of a recurrent structural motif. To investigate further we take the geometry of the the stable figure eight and use it as a starting point to construct new trial initialisation geometries. We do so in the simplest way possible --- as highlighted in the dotted circle around a section of the $T=0$ figure eight in figure \ref{fig:StableKnots}, the knot geometry contains a half-turn of a helix, which we may extend to an integer number of half-turns. Doing so gives a family of trial initialisation curves alternating between knots and two component links, the next two being the Whitehead link and the $6_2$ knot. Simulation reveals that such initialisation geometries evolve to apparently stable states. Figure \ref{fig:StableKnots} shows their detailed geometry, and in figure \ref{fig:Whitehead_6_2} we confirm their bulk stability up to $T=15000$. Both states share the twofold symmetry and tight synchronisation over a vortex rotation period found in the $3_{1,2}$ and figure eight knots, with especially close similarity in the geometry and twist distributions of the $4_1$, Whitehead link and $6_2$ knots. This similarity suggests that they arise as the start of a family of such stable knots which does not cleave along some existing sub-category of knots (for example torus knots) but rather arises specifically from the FitzHugh-Nagumo dynamics. As another demonstration of the importance of initial conditions, and a reminder that such states may have small basins of attraction, we note that the $6_2$ of \S \ref{sec:UnstableKnots} does not find this stable state over the times simulated.

\begin{figure*}[htbp]
    \includegraphics[width=0.8\textwidth]{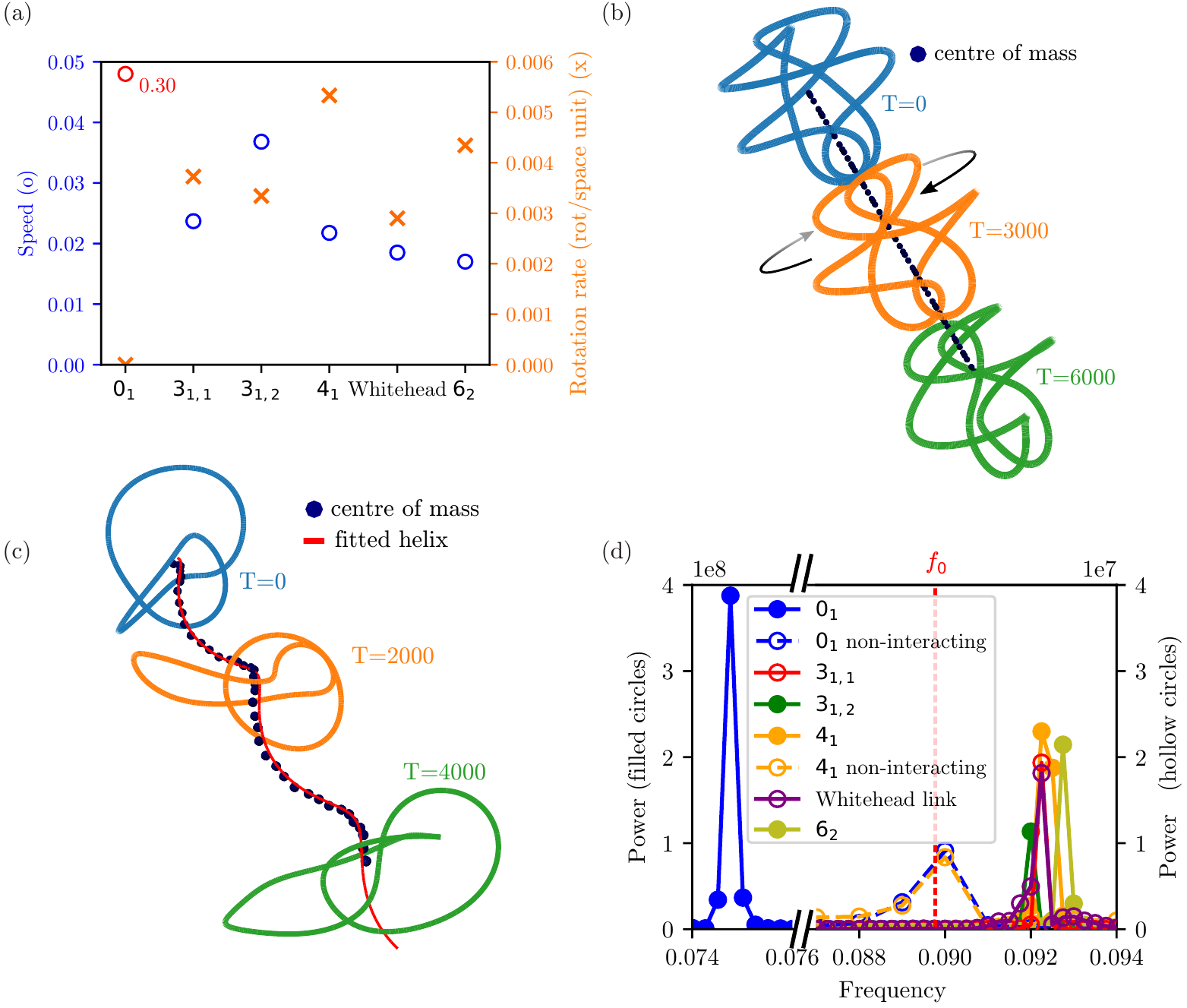}
    \caption{Dynamics of stable knots. (a) A summary of drift speeds and rotation rates for all known stable knots. Note that the velocity given for the $3_{1,1}$ is that along its helical axis. (b) Drift and rotation of the $6_2$ knot. Shown are centres of mass taken at $T=100$ intervals (blue dots), and snapshots of the geometry at $T=3000$ intervals; between each snapshot the knot has rotated $\sim 20$ times. (c) The $3_{1,1}$ drifts along a helical path (fitted red curve), rotating about the helix axis as a rigid body. (d) Power spectra of high frequency oscillations in knot length data. Before inter-vortex interactions occur, the oscillation period is the same as ${f_0}$. All non-trivial stable knots show a similar shift to higher vortex rotation frequencies ($T=0.97T_0$), with the unknot alone showing a shift to lower frequencies ($T=1.19T_0$).}
\label{fig:KnotDynamics}
\end{figure*}

In figure \ref{fig:KnotDynamics} we explore the dynamical properties of all stable knots found thus far. With the exception of the $3_{1,1}$, we find that each drifts along its axis of symmetry and rotates as a rigid body, with speeds and rotation rates summarised in figure \ref{fig:KnotDynamics}(a); these rates are computed by averaging the motion of the rigid body frame of the stable knot over $\Delta T = 2000$. An example of this motion is shown for the $6_2$ knot in figure \ref{fig:KnotDynamics}(b) (as can be seen in figure \ref{fig:Whitehead_6_2}, the Whitehead link and the $6_2$ knot have some long timescale periodic length modulation which corresponds to a slight oscillation in their velocity). The $3_{1,1}$ instead drifts in a helix as shown in figure~\ref{fig:KnotDynamics}(c), rotating about the helical axis as a rigid body, a reflection of its lack of threefold symmetry (a numerical fit to this helix \cite{Enkhbayar2008} finds that it has radius $3.23$ and pitch $39.34$). The scale, and structure with knot size, of the drift velocities resembles that found for torus links in Ref.~\cite{Maucher2018b}: within the family of knots discussed above, we see drift velocity decreasing with knot size. However, the complex geometries of the stable knots discussed here renders the explanation for this decrease given for torus knots (decreasing asymmetry between inner and outer parts of the torus as size increases) inapplicable. A reflection of this complexity is that, beyond consistency of scale, there is no clear accompanying pattern in the rotation rate data.

We briefly note that in the above discussion of vortex rotation sense, drift velocity and overall knot rotation sense we have not been careful to distinguish the possibly different behaviours of oriented or chiral variants from one another. All stable knots and links discussed above are isotopic to themselves under reversal of the orientation of any link component, however with the exception of the $4_1$ they are all chiral, and this chirality determines the rotation sense of the knot. In figures \ref{fig:StableKnots} and \ref{fig:KnotDynamics} we present variants rotating in a right handed sense about their drift velocity; left handed variants, with reversed twist distributions, also exist.

As discussed in \S \ref{sec:UnstableKnots}, Ref.~\cite{Maucher2018a} reports an increase in the rotation period of a stable unknot by $14\%$. We investigate whether similar shifts exist for other stable knots by looking at the spectra of their high frequency length oscillations. Figure~\ref{fig:KnotDynamics}(d) shows the spectra of all stable knots as measured over $\Delta T = 4000$ after they reach their stable configurations, alongside the spectra of the first $T=1000$ of the unknot and $4_1$ data shown in figure~\ref{fig:UnstableKnotsLength}. We include this second set of data for calibration and methodology validation, as during this time we expect the data to give the spectrum of a noninteracting knot, which should approximately correspond to $f_0$. As expected, the length oscillations of the noninteracting data are consistent with the fundamental vortex rotation frequency of $f_0=0.0898$, and do not vary with knot topology --- before inter-vortex interactions occur the global structure of the filament does not dramatically affect vortex rotation period. In fact, as this data is taken during the contraction of both knots, the observation that it shows purely spectral broadening suggests a negligible role for curvature in possible shifts to rotation frequency. As with the noninteracting data, the stable knot spectra show single peaks, but their frequencies are shifted relative to the noninteracting case on a scale which exceeds our estimate of curvature induced corrections. For all nontrivial knots, this shift is to a higher frequency (lower period), and its size is approximately constant; we obtain a period of $T=0.97T_0$. By contrast, the unknot alone shows a substantial shift to lower frequencies (higher period); we find an unknot rotation period of $T=1.19T_0$, consistent with the results of Ref.~\cite{Maucher2018a}. That the situation for nontrivial knots is a shift to lower period relative to an isolated filament is intriguing, and suggests itself as a potential origin of the motion of the collision interface leading to the wave slapping observed in \S \ref{sec:UnstableKnots}. One important complicating factor in this sort of analysis is Doppler shift. Although the period of the stable unknot is higher than $T_0$, its velocity is $0.3$, a substantial fraction of the wavespeed $(1.9)$ in the medium. Using the data presented here this gives a Doppler shifted period for a stationary observer ahead of the unknot of only $1\%$ greater than $T_0$; in other words, at least for the unknot, relative filament motion is extremely important in determining the stability of a situation. A similar calculation for an observer behind the $4_1$ gives a Doppler shifted period of $T=0.975T_0$, a far less substantial shift. We speculate that these two facts, firstly that stable structures generically appear to have periods shifted below $T_0$, and secondly that even when the shift is to a higher period in the case of the unknot (extrapolating unknot behaviour to that of generic anti-parallel strands) this increase is compensated for in a directional manner by Doppler shift, give an intrinsically unstable dynamics in which the formation of any interacting structure hinders further formation via wave slapping.

\section{\label{sec:Discussion}Discussion}

We have presented a survey of the bulk dynamics of knotted vortices in the FitzHugh-Nagumo model covering prime knots up to crossing number $N=8$. Although the simplest knots --- the unknot, trefoil and figure eight --- possess stable states and exhibit a fascinating dynamics of untangling without reconnections, this is not repeated for any of the other knots in our survey. The general trend is an irregular dynamics, marked by sustained periods of length expansion of parts of the knot, the cause of which we have directly shown to be a long-range wave slapping interaction. In several cases, this wave slapping led to strand reconnections and topology change, phenomena which appear to be associated more with the geometry of the wavefield than the topology of the vortex. 

For those stable knots found in our initial survey, we have tested the effectiveness of the FitzHugh-Nagumo flow in untangling a wide variety of initial conditions. Although the dynamics successfully untangled all but one initial geometry over the time simulated, we saw that in the case of the figure eight knot this untangling was far from monotonic, and that the same wave slapping dominating high crossing number knot behaviour may also cause low crossing number knots to substantially increase in length before untangling. These results stand in contrast to those of Ref.~\cite{Maucher2016} and our own on the rapid untangling of unknots. We gave a detailed characterisation of the geometry and dynamics of all known stable vortices in the bulk, including the tight synchronisation of their associated wavefields, their motion through the medium and natural rotation periods, and shifts in the spectra of their high frequency length oscillations. In addition to the already known trefoil and figure eight knots, we found stable forms for the Whitehead link and $6_2$ knot, both of which appear to come from the same `family' of knots as the figure eight. While in the former case, the basin of attraction appears to be large, the same cannot be said for the latter, at least for the timescales of the simulations we have run. 

Throughout this paper, we have emphasised the importance of the collision interface on understanding long timescale vortex dynamics. Although we have seen many examples of its importance, an understanding of its own dynamics is currently qualitative at best. As a first step towards rectifying this, it would be interesting to directly study the evolution of local rotation rate along a vortex to fully disentangle the possible effects of curvature, twist and interactions. Turning from general dynamical questions to the details of stable states, beyond noting close similarities between those found we have not proposed principles by which their geometry and behaviour may be understood. A detailed description appears challenging, but the observed wavefield synchronisation, and similarities in the size of spectral shifts, of \S\ref{sec:StableKnots} offers a global organising principle from which one might try to predict geometries. When discussing stability we have contrasted our own results in the bulk with those on torus knots and links that have been found near no-flux boundaries~\cite{Maucher2017,Maucher2018b}, detailing where results overlap (the stability of the trefoil and figure eight) and where they diverge. Although we have seen that boundary stabilisation is more complex than simply a suppression of the sproing instability, its exact nature remains unclear, and deserves further study. 

The features of the FitzHugh-Nagumo model at the parameter values studied here which are conducive to the formation of stable knots --- short-range inter-vortex repulsion, a contractile filament law of motion --- are offset by other undesirable features, primarily wave slapping. Parameter choices were originally made in Ref.~\cite{Henze1993} on the basis that such values gave two-dimensional vortices with desirable properties and three-dimensional simulations were computationally feasible. It would be interesting to revisit these choices armed with new criteria for a desirable set of parameters. For example, we might search for parameters (or indeed models) such that rotation frequency is seen to decrease with twist and interactions. A related question is to explore whether wave slapping interactions have any role in enhancing untangling as well as hindering it --- in other words, whether the untangling aspect of the dynamics can be captured in a local geometric model. Here the tubed knot of \S\ref{sec:UnstableKnots} offers some hints; in preliminary simulations of tubed versions of the tangled unknots of \S\ref{sec:StableKnots} we do not see substantial differences in the untangling times between tubed and untubed unknots. It may be the case that, although the full dynamics appear extremely difficult to capture with a local geometric model, such a model offers insight for the restricted case of unknot untangling. This is especially interesting given the apparent contrast between the untangling dynamics seen here and those utilised by line tension minimisation methods \cite{Maucher2016}. 

Stable vortex rings have been realised experimentally and successfully described using existing theory \cite{Steinbock2006, Azhand2014, Totz2015}. As such, although one expects the precise details of knot stability to be specific to the system studied, we believe that our exploration of the phenomena seen here --- the importance of wave slapping, bulk simplification of low crossing number knots, frequency shifts in stable knots --- is of direct experimental interest for a general excitable medium, outside of the details of the FitzHugh-Nagumo model. 

\begin{acknowledgments}
{We are grateful to P. Sutcliffe for useful discussions. JB acknowledges partial support from a David Crighton Fellowship and thanks R.E.~Goldstein for his discussions on spectral methods, as well as his and DAMTP's hospitality during the fellowship. This work was supported by the UK EPSRC through Grant Nos. EP/L015374/1 (JB) and EP/N007883/1 (CAW and GPA). }
\end{acknowledgments}

\end{document}